\documentclass[journal]{IEEEtran}

\usepackage{threeparttable}
\usepackage{cite}
\usepackage{amsmath,amssymb,amsfonts}
\usepackage{algorithmic}
\usepackage{graphicx,color}
\usepackage{textcomp}
\usepackage{xcolor}
\usepackage{hyperref}

\usepackage{lipsum} 
\usepackage{multirow}
\usepackage{comment}

\usepackage[printonlyused,withpage]{acronym}
\newacro{5G}{Fifth Generation Wireless Specifications}
\newacro{ACK}{acknowledgment}
\newacro{AP}{Application Providers}
\newacro{AF}{Application Function}
\newacro{BS}{Base Station}
\newacro{API}{Application Program Interface}
\newacro{AR}{Augmented Reality}
\newacro{ARQ}{Automatic Repeat Query}
\newacro{AWS}{Amazon Web Services}
\newacro{BER}{Bit Error Rate}
\newacro{CAPEX}{Capital Expenditure}
\newacro{CDN}{Content Delivery Network}
\newacro{CDF}{Cumulative Distribution Function}
\newacro{CLI}{Command Line Interface}
\newacro{CPS}{Cyber-Physical System}
\newacro{CPU}{Central Processing Unit}
\newacro{CRC}{Cyclic Redundancy Check}
\newacro{CSI}{Channel State Information}
\newacro{DANN}{domain adversarial neural networks}
\newacro{DB}{Database}
\newacro{DC}{Data Center}
\newacro{DRL}{Deep Reinforcement Learning}
\newacro{DoF}{Degrees of Freedom}
\newacro{DOF}{Degrees Of Freedom}
\newacro{EC2}{Elastic Compute Cloud}
\newacro{FaaS}{Function-as-a-Service}
\newacro{FEC}{Forward Error Correction}
\newacro{FPGA}{Field-Programmable Gate Array}
\newacro{GUI}{Graphical User Interface}
\newacro{HARQ}{Hybrid Automatic Repeat Query}
\newacro{HW}{Hardware}
\newacro{IaaS}{Infrastructure as a Service}
\newacro{i.i.d}{Independent and Identically Distributed random variables}
\newacro{IoE}{Internet of Everything}
\newacro{I/O}{Input/Output}
\newacro{IoT}{Internet of Things}
\newacro{IP}{Infrastructure Providers}
\newacro{LDPC}{Low-Density Parity-Check }
\newacro{LTE}{Long Term Evolution}
\newacro{LSTM}{Long Short-Term Memory}
\newacro{LuMaMi}{Lund Massive MIMO}
\newacro{MAC}{Medium Access Control}
\newacro{MAN}{Metropolitan Area Network}
\newacro{MAE}{mean absolute error}
\newacro{MAPE}{mean absolute percentage error}
\newacro{MSE}{mean square error}
\newacro{MCN}{Heterogeneous Distributed Computing}
\newacro{MCN}{Mobile Cloud Network}
\newacro{MD}{Mobile Device}
\newacro{MD}{Mobile Devices}
\newacro{MEC}{Mobile Edge cloud}
\newacro{MIMO}{Multiple Inputs Multiple Outputs}
\newacro{mimo}[MIMO]{multiple-input-multiple-output}
\newacro{MLP}{multilayer perceptron}
\newacro{NWDAF}{Network Data Analytics Function}
\newacro{NEF}{Network Exposure Function}
\newacro{SISO}{Single Input Single Output}
\newacro{siso}[SISO]{single-input-single-output}
\newacro{MIP}{Mixed Integer Programming}
\newacro{mMTC}{massive Machine Type Communication}
\newacro{MN}{Mobile Network}
\newacro{MNO}{Mobile Network Operator}
\newacro{MPC}{Model Predictive Controller}
\newacro{MQTT}{Message Queue Telemetry Transport}
\newacro{MR}{Maximum-Ration Combining}
\newacro{MT}{Mobile Terminal}
\newacro{MTU}{Maximum Transmission Unit}
\newacro{MU-MIMO}{Multi-User MIMO}
\newacro{NFV}{Network Function Virtualisation}
\newacro{NoOps}{No Operations}
\newacro{OFDM}{Orthogonal Frequency-Division Multiplexing}
\newacro{OPEX}{Operational Expenditure}
\newacro{PaaS}{Platform as a Service}
\newacro{PDC}{Proximal Data Centers}
\newacro{PID}{Proportional Integral Derivative}
\newacro{PM}{Physical Machine}
\newacro{QoS}{Quality of Service}
\newacro{QPSK}{Quadrature Phase Shift Keying}
\newacro{RAN}{Radio Access Network}
\newacro{RAT}{Radio Access Technology}
\newacro{RBS}{Radio Base Station}
\newacro{RBS}{Radio Base Stations}
\newacro{RDC}{Remote Data Centers}
\newacro{RTD}{Round-Trip Delay time}
\newacro{RTT}{Round Trip Time}
\newacro{OWD}{One-Way Delay}
\newacro{SaaS}{Software as a Service}
\newacro{SDK}{Software Development Kit}
\newacro{SDN}{Software Defined Networks}
\newacro{SDR}{Software Defined Radio}
\newacro{SLA}{Service Level Agreement}
\newacro{SLO}{Service Level Objectives}
\newacro{slo}{service level objectives}
\newacro{SNR}{Signal-to-Interference-plus-Noise Ratio}
\newacro{SINR}{Signal-to-Interference-plus-Noise Ratio}
\newacro{CSI}{Channel State Information}
\newacro{CQI}{Channel Quality Indicator}
\newacro{SNS}{Simple Notification Service}
\newacro{SoS}{System of Systems}
\newacro{SP}{Service Providers}
\newacro{SQL}{Structured Query Language}
\newacro{SUMO}{Simulation of Urban MObility}
\newacro{SW}{Software}
\newacro{TLS}{Transport Layer Security}
\newacro{TraCI}{Traffic Control Interface}
\newacro{TSC}{Traffic Signal Control}
\newacro{TSP}{Transit Signal Priority}
\newacro{TTI}{Transmission Time Interval}
\newacro{TWAMP}{Two-Way Active Measurement Protocol}
\newacro{UE}{User Equipment}
\newacro{UM}{User Mobility}
\newacro{URLLC}{Ultra-Reliable and Low-Latency Communication}
\newacro{UX}{User Experience}
\newacro{WAN}{Wide Area Network}
\newacro{WLAN}{Wireless Local Area Network}
\newacro{VM}{Virtual Machine}
\newacro{WSN}{Wireless Sensor Network}
\newacro{vSoftPLC}{virtual Software Programmable Logic Controller}
\newacro{PLC}{Programmable Logic Controller}
\newacro{ZF}{Zero-Forcing}
\newacro{MS}{Mobile Station}
\newacro{NTP}{Network Time Protocol}
\newacro{PTP}{Precision Time Protocol}
\newacro{NGCC}{Next Generation Cloud Computing}
\newacro{ERDC}{Ericsson Research Data Center}
\newacro{DNR}{Distributed-NodeRED}
\newacro{ADC}{Analog to Digital Converter}
\newacro{DAC}{Digital to Analog Converter}
\newacro{NoOps}{No-Operations}
\newacro{PaaS}{Platform-as-a-Service}
\newacro{WASP}{Wallenberg Autonoms Systems and Software Program}
\newacro{COTS}{Commercial off-the-shelf}
\newacro{COTC}[CotC]{Control over the Cloud}
\newacro{PoC}{Proof of concept}
\newacro{B'n'B}{Ball and beam}
\newacro{LQR}{Linear–Quadratic regulator}
\newacro{LQ}{Linear Quadratic}
\newacro{K8S}{Kubernetes}
\newacro{OSI}{Open Systems Interconnection}
\newacro{IIoT}{Industrial Internet-of-things}
\newacro{REST}{Representational state transfer}
\newacro{ReLU}{Rectified Linear Unit}
\newacro{ICMP}{Internet Control Message Protocol}
\newacro{UDP}{User Datagram Protocol}
\newacro{TCP}{Transmission Control Protocol}
\newacro{CGI}{Common Gateway Interface}
\newacro{LAN}{Local Area Network}
\newacro{IT}{Information Technology}
\newacro{HTTP}{Hypertext Transfer Protocol}
\newacro{HA}{High availability}
\newacro{RAV}{Relative Accumulated Violations}
\newacro{RAE}{Relative Accumulated Error}
\newacro{RMCV}{Relative Maximum Constraint Violation}
\newacro{EDC}{Edge Data Center}
\newacro{TSN}{Time Sensitive Networks}
\newacro{ICT}{Information and Communications Technology}
\newacro{CCS}{Cloud Control System}
\newacro{NCS}{Networked Control System}
\newacro{MLE}{Maximum Likelihood Estimation}
\newacro{R-CCS}{Resilient Cloud Control System}
\newacro{ML}{Machine Learning}
\newacro{NSA}{non-standalone}
\newacro{eNB}{eNodeB}
\newacro{gNB}{gNodeB}
\newacro{NR}{New Radio}
\newacro{TDD}{Time-Division Duplex}
\newacro{DL}{downlink}
\newacro{UL}{uplink}
\newacro{UPF}{User Plane Function}
\newacro{V2X}{Vehicle-to-everything}
\newacro{CAV}{Connected and Automated Vehicle}
\newacro{ToD}{Tele-operated driving}

\usepackage{siunitx}
\DeclareSIUnit\Mbps{Mbps}
\DeclareSIUnit{\nothing}{\relax}

\hypersetup{hidelinks}
\usepackage{algorithm,algorithmic}
\def\BibTeX{{\rm B\kern-.05em{\sc i\kern-.025em b}\kern-.08em
    T\kern-.1667em\lower.7ex\hbox{E}\kern-.125emX}}
\AtBeginDocument{\definecolor{tmlcncolor}{cmyk}{0.93,0.59,0.15,0.02}\definecolor{NavyBlue}{RGB}{0,86,125}}

\begin{document}


\title{Multi-Generator Continual Learning for Robust Delay Prediction in 6G

\author{Xiaoyu Lan, Jalil Taghia, Hannes Larsson, Andreas Johnsson}

\thanks{All the authors are with Ericsson Research, Ericsson AB, Stockholm, Sweden.}
\thanks{This paper has been accepted for publication in IEEE Transactions on Machine Learning in Communications and Networking.\\
DOI: 10.1109/TMLCN.2026.3663092\\
https://doi.org/10.1109/TMLCN.2026.3663092}}

\maketitle

\begin{abstract}
In future 6G networks, dependable networks will enable telecommunication services such as remote control of robots or vehicles with strict requirements on end-to-end network performance in terms of delay, delay variation, tail distributions, and throughput. With respect to such networks, it is paramount to be able to determine what performance level the network segment can guarantee at a given point in time. One promising approach is to use predictive models trained using machine learning (ML). Predicting performance metrics such as one-way delay (OWD), in a timely manner, provides valuable insights for the network, user equipments (UEs), and applications to address performance trends, deviations, and violations. Over the course of time, a dynamic network environment results in distributional shifts, which causes catastrophic forgetting and drop of ML model performance. In continual learning (CL), the model aims to achieve a balance between stability and plasticity, enabling new information to be learned while preserving previously learned knowledge. In this paper, we target on the challenges of catastrophic forgetting of OWD prediction model. We propose a novel approach which introducing the concept of multi-generator for the state-of-the-art CL generative replay framework, along with tabular variational autoencoders (TVAE) as generators. The domain knowledge of UE capabilities is incorporated into the learning process for determining generator setup and relevance. The proposed approach is evaluated across a diverse set of scenarios with data that is collected in a realistic 5G testbed, demonstrating its outstanding performance in comparison to baselines. 

\end{abstract}


\section{INTRODUCTION}
\label{sec:introduction}
\IEEEPARstart{T}{he} evolution towards 6G is expected to enable a new generation of applications, for example in the area of cyber-physical systems \cite{Karapantelakis2024, liu2017review}, demanding ultra-reliable, real-time, and low-latency communication. For such applications, it is crucial to assess performance indicators such as \ac{OWD} \cite{owd} and \ac{RTT} \cite{rtt}, as even minor disturbances in the communication between two network functions could lead to severe operational and safety risks. Unfortunately, there is complexity associated with assessing the performance outcome of a \ac{UE} as it is highly influenced by and dependent on multiple factors such as network configuration, signal conditions, competing traffic, exogenous processes in the \ac{UE}, and variations in \ac{UE} hardware \cite{ansari2022performance}. 
A promising approach, explored in both academia and industry, is based upon \ac{ML} where the performance outcome of a \ac{UE}, such as the \ac{OWD}~\cite{riihijarvi2018machine, mostafavi2023data, rao2022prediction}, is predicted based upon statistics available to the network operator. Accurate performance prediction, not only with respect to point estimates but also with respect to distributions and tails \cite{mostafavi2023data, flinta2020predicting, samani2021conditional}, is essential for proactive service assurance, troubleshooting, verification, and performance optimization. 

\begin{figure*}
\centerline{\includegraphics[width=5.3in]{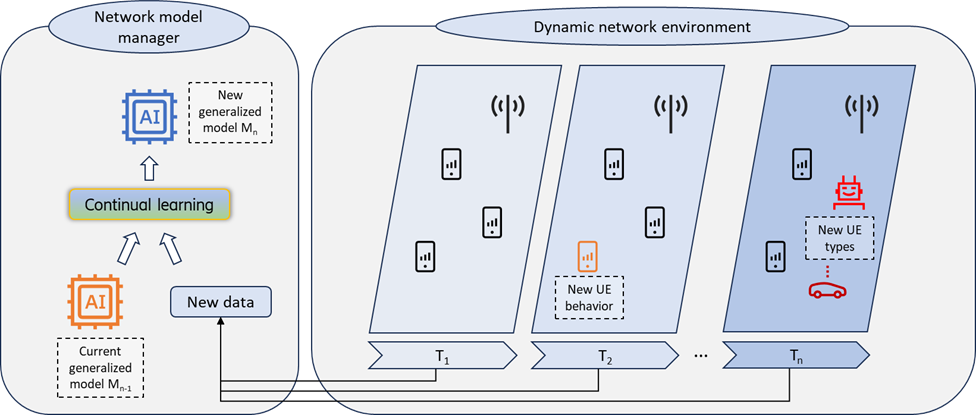}}
\caption{A dynamic network environment gives rise to distributional shifts in the data, causing loss of \ac{OWD} model performance. Continual learning can balance model plasticity and stability, ensuring new information is captured in the model while preserving old knowledge, that is avoiding catastrophic forgetting. }
\label{fig:scenario}
\end{figure*}

Previous research has demonstrated the feasibility of modeling delay; however, it has also shown that model performance deteriorates over time due to distributional shifts in data. These shifts are due to factors such as previously unseen user behavior patterns (e.g., traffic fluctuations, mobility) and variations in \ac{UE} chip sets \cite{rao2024generalizable}. Various mitigation strategies have been explored, including transfer learning and domain adaptation \cite{larsson2023domain}, and semi-supervised learning \cite{rao2025self}. More broadly, it is well established that network management models degrade over time due to the dynamic nature of networks and their underlying compute resources \cite{moradi2019performance}. The challenge of maintaining network management models over time has also been identified in standardization, and thus there is an increased interest in incorporating automated \ac{ML} workflows and continuous operations into the network architecture. 

Despite recent advancements in model generalization, a significant challenge remains: sustaining model performance over extended periods. This challenge is illustrated in Fig.~\ref{fig:scenario}, where a network environment evolves over time, introducing distributional shifts that can degrade model performance. Each time step in this dynamic environment can be considered as a continual learning (CL) task, new data is observed and thereafter integrated with the existing generalized model through CL, producing an updated generalized model. The key challenge lies in balancing model plasticity with learning stability \cite{dohare2024loss} - ensuring that new behaviors in the network environment are learned while preserving previously acquired knowledge. In other words, preventing catastrophic forgetting in ML \cite{McCloskey1989CatastrophicII}.
This paper specifically addresses this challenge for \ac{OWD} prediction models, as discussed above, and proposes a novel approach based on CL. The method leverages generative replay \cite{Shin2017ContinualLW} and extends such approaches with a multi-generator framework, where generators are designed and maintained using domain knowledge of the network and its \ac{UE}s. Generative replay, specifically the proposed multi-generator approach, not only mitigates catastrophic forgetting, it also improves performance on the distribution tail as well as enabling network operators to reduce the storage requirements for previously seen data samples.

The main contributions of this paper are as follows. We propose a novel method for CL based on generative replay, introducing the concept of multi-generator in combination with tabular variational autoencoders (TVAE) to mitigate catastrophic forgetting in \ac{OWD} prediction.  Additionally, we elaborate on a method for determining generator setup and relevance, leveraging knowledge of \ac{UE} capabilities, thereby incorporating domain knowledge into the learning process. The proposed approach is evaluated across a wide range of scenarios in a realistic 5G testbed, demonstrating its effectiveness in comparison to a baseline both from a point estimate perspective, and also on the tail. Moreover, we highlight the reduced need for data storage, targeting the challenges of resource constraints in 5G networks, which enhances the efficiency of the method.

The rest of the paper is organized as follows. Section \ref{sec:background} describes the background on our use case and problem formulation. Section \ref{sec:approach} presents the necessary background on deep generative replay and introduces the novel approach on multi-generator CL, whereas Section \ref{sec:dataset} describes our testbed and datasets. Section \ref{sec:evaluation} describes an in-depth evaluation framework of the approach and Section \ref{sec:results} presents the evaluation results. In Section \ref{sec:discussion}, we provide discussions related to the approach. 
Section \ref{sec:relatedwork} contains related work, and conclusions are found in Section \ref{sec:conclusions}.


\section{PROBLEM DEFINITION}  
\label{sec:background}
In this section, we provide a high-level background on \ac{OWD} prediction, proceed with notations and framework of CL, and present problem statement.

\subsection{BACKGROUND}

The objective of this paper is to establish a robust approach for training and maintaining a \ac{ML} model that predicts \ac{OWD} as experienced by a \ac{UE}. This prediction is based on measurements taken with high frequency from the baseband in the \ac{RAN}. Although our previous work \cite{rao2024generalizable} addressed some aspects of \ac{OWD} model generalization, the expanded aim of this paper is to prevent catastrophic forgetting, and thus ensure robustness during long-term maintenance of \ac{OWD} models.

In the scenario illustrated in Fig.~\ref{fig:scenario}, a model manager is tasked with maintaining a \ac{OWD} model, which was originally trained using historical data. These data distributions are influenced by a variety of factors, such as radio conditions, network configurations, network load, \ac{UE} movement patterns, and \ac{UE} device types. As the network evolves over time, the resulting dynamicity leads to  distributional shifts in the measurements; among others, the dynamicity is related to new \ac{UE} behavior, previously unseen \ac{UE} types, and potentially altered network configurations. Consequently, in the subsequent time slot, the \ac{OWD} and baseband features exhibit new distributions.

The challenge targeted in this paper is to maintain the ML model performance over time, leveraging and extending the concept of CL for retraining the model with new data while preserving knowledge related to previous steps.

\subsection{NOTATIONS AND FRAMEWORK}
In the following, we describe the notations and framework. Here, the task is prediction of \ac{OWD} 
as observed by the \ac{UE} using \ac{RAN} data collected in the baseband. 

We denote the ML task by $T$. The OWD measurements from the \ac{UE} are denoted by $y$ distributed according to $P(Y)$. Measurements from the \ac{RAN} are referred to as the features. Let ${x_j}$ denote the $j$-th feature distributed according to ${P(X_j)}$. A collection of $d$ different features are shown for convenience in the vector notation by ${\boldsymbol x\in \mathbb R^{d}}$; however, note that no explicit spatial information is assumed across the elements of $\boldsymbol x$, indicating that we are concerned with tabular data.

We assume a CL scenario where the underlying ML task, $T$, remains the same. However, the task condition may change due to the effect of external factors that result in distributional shift in space of $X$ and $Y$. In the context of \ac{OWD} prediction, example of such external factors are the \ac{UE}'s device type along with the UE position and movement patterns (refer to Table~\ref{tb:varied-param}). Let ${T^{(i)}}$ denote the task at the $i$-th condition. A set of such task conditions are denoted by ${\mathcal T=\{T^{(i)}\mid i=1,2,\ldots, I\}}$, where $I$ is the number of tasks. In CL framework, the set $\mathcal T$ is treated as a sequence. 

Let ${\mathcal D_{T^{(i)}}=\{\boldsymbol x_n^{(i)}, y_n^{(i)} \mid n=1, \ldots, N^{(i)}\}}$ denote our dataset consisting of pairs of input features and output targets at the $i$-th task condition $T^{(i)}$, where $N$ is the number of samples in the dataset. Further, let $M$ denote a ML model that serves as the predictive model of $y$ given $\boldsymbol x$. In the CL framework, task conditions are introduced sequentially which points at the sequential learning of the predictive model. 

Let $\theta$ denote the parameter set of the predictive model. The learning begins with the dataset from a given task condition, ${M\left(\mathcal D_{T^{(i)}}; \theta \right)}$ parameterized by $\theta$, and follows with the next task condition, ${M\left(\mathcal D_{T^{(i+1)}}; \theta \right)}$. However, a major complication is that the datasets from two arbitrary task conditions, $\mathcal D_{T^{(i)}}$ and $\mathcal D_{T^{(i+1)}}$, may not share the same underlying distributions, that is ${P(X_j^{(i)}){\neq}P(X_j^{(i+1)}) }$ for the $j$-th feature and ${P(y^{(i)}){\neq}P(y^{(i+1)})}$. As the result, the learned model would have a reduced predictive relevance for data from the $i$-th task condition. The problem is commonly referred to as the \emph{catastrophic forgetting} which magnifies as sequential learning continues.

One line of approaches for solving the catastrophic forgetting problem is to refrain the model from forgetting by presenting data, or a subset of data, from the past task conditions, referred to as the \emph{data replay}. While highly effective, the assumption of having access to the past data, or a subset of which, is restrictive for many practical applications, especially in resource-constrained telecom network environments, due to the cost of data storage as well as privacy concerns.

\subsection{PROBLEM STATEMENT}
The \emph{deep generative replay} framework \cite{Shin2017ContinualLW} presents a promising approach to CL, addressing the limitations associated with the data replay. While potentially effective, \emph{we argue that the generative replay itself can suffer from catastrophic forgetting}. As the number of tasks increases over time, with the growing diversity resulting from introduction of new tasks, the generative model tends to capture merely the modalities in the data that represent the bulk of the underlying data distribution and fail to capture all the modalities. We refer to this phenomenon as \emph{mode collapse} that is when a modality that was present once collapses into a more dominant modality. 

The main reason for the mode collapse is the limited expressiveness capability of the underlying generative model. One way to improve the model expressiveness is to choose a generative model suitable for the data under consideration. We show that this has a major influence in the overall success of the generative replay in the framework of CL. Specifically, in the context of OWD prediction that is concerned with tabular data, we show that using a generative model tailored for tabular data substantially improves the performance. However, this does not resolve the mode collapse if the data generation process differs considerably across tasks. As an example, the \ac{UE} device type is a factor that gives rise to different data generation processes. In this regard, we argue that having multiple generative models where each represents one of the modalities of their respective data generation processes can further improve the overall model expressiveness. However, the question is how one would select the relevant generative model from a given collection of generative models? This points to a need for a generator selector. One approach is to construct the generator selector in a data-driven fashion. Alternatively, one can leverage domain knowledge in construction of the generator selector. As an example, in the context of our running example, understanding of the \ac{UE} device type is knowledge that can help select the right generator. Here we show that incorporating domain knowledge can significantly mitigate catastrophic forgetting by providing contextual frameworks that guide learning processes. By embedding these domain-specific insights into the learning architecture, models can develop more stable representations that reduce the risk of mode collapse (or in other words catastrophic forgetting) as a new task arises. 

In summary, in the context of \ac{OWD} prediction, we make the following contributions to reduce risks of catastrophic forgetting: (1) use of a generative model that is tailored for tabular data, (2) construction of a multi-generator architecture for the generative replay, and (3) devising a domain-guided generator selector. We show via empirical experiments that the proposed steps substantially reduce risks of catastrophic forgetting and contribute to improving model's ability to maintain performance across diverse tasks.

\section{MULTI-GENERATOR CONTINUAL LEARNING}  
\label{sec:approach}

In this section, we describe our novel domain-guided approach to construction of a multi-generator-based generative replay in the framework of CL. We start this section with a background of the preliminaries needed for our construction and proceed with the proposed approach.

\subsection{BACKGROUND}
\subsubsection{Deep generative replay}
The model architecture of the deep generative replay consists of a generative model, \emph{generator}, and a task solving model, \emph{solver}. Shin et al. in \cite{Shin2017ContinualLW} refer to this dual model architecture as \emph{scholar}, defined by ${\mathcal H=\{G, S\}}$ where $G$ denotes the generator and $S$ denotes the solver. The generator is a parameterized generative model, with a parameter set denoted by $\phi$, that is learned to produce real-like samples of the input features, and the solver is a parameterized predictive model of the task with a parameter set denoted by $\theta$. 

The components of the scholar, generator and solver, are learned in a two-step sequential-learning framework. Let ${\mathcal H^{(i)}}$ denote the scholar model at the $i$-th task condition. The generator and the solver of ${\mathcal H^{(i)}}$ are denoted by ${G(\phi^{(i)})}$ and ${S(\theta^{(i)})}$. Furthermore, for convenience, we use the following notations equivalently: ${G^{(i)}\equiv G(\phi^{(i)})}$ and similarly,  ${S^{(i)}\equiv S(\theta^{(i)})}$.

The training procedure involved in learning of the current scholar, ${\mathcal H^{(i)}}$, from the prior scholar, ${\mathcal H^{(i-1)}}$, involves two \emph{independent} training procedures for the generator and the solver. 

In the first step, from the prior state of the generator, ${G^{(i-1)}}$, a set of artificial input data samples are produced, referred to as the replay input features and denoted by ${\boldsymbol x^\prime}$. The replay targets, ${ y^\prime}$, are predicted from the prior state of the solver, ${S^{(i-1)}}$, given replay inputs, $\boldsymbol x^\prime$, according to: ${ y^\prime = S\left(\boldsymbol x^\prime; \theta^{(i-1)}\right)}$, where ${\theta^{(i-1)}}$ denotes the prior state of the solver's parameter set. The replay targets are the solver's response to the replay input in the past. Finally, given the input features from the current task condition, ${\boldsymbol x}$, the targets, $y$, and the replay targets,  $y^\prime$, a training loss is constructed according to:

\begin{multline}
\label{eq:loss_sol1}
\ell_{S} = \alpha \mathbb E_{(\boldsymbol x, y)\sim \mathcal D_{T^{(i)}}}\left[ L_S\left(S\left(\boldsymbol x; \theta^{(i)}\right), y \right)\right] \\ + (1-\alpha) \mathbb E_{\boldsymbol x^\prime\sim G^{(i-1)}}\left[ L_{S}\left( S\left(\boldsymbol x^\prime; \theta^{(i)} \right),  y^\prime \right)  \right],
\end{multline}
where ${\boldsymbol x^\prime \sim G^{(i-1)}}$ denotes the replay inputs drawn from the prior generator, ${(\boldsymbol x, y)\sim \mathcal D_{T^{(i)}}}$ denotes the pair of input samples and targets taken randomly from the dataset of the $i$-th task condition, $\alpha$ is a ratio of mixing real data from the current task condition with real-like data from the previous task condition and $L_S$ denotes the solver's loss function. This step updates the solver of the $i$-th scholar.

Independent from the first step, in the second step, the replay inputs $\boldsymbol x^\prime$ are mixed with the input features from the current task condition, ${\boldsymbol x}$. Let ${\widetilde{\boldsymbol x}:=(\boldsymbol x, \boldsymbol x^\prime)}$ define the concatenation of the replay inputs with the current input features. Given ${\widetilde{\boldsymbol x}}$, the generator is trained (or adapted) to learn to generate samples from their cumulative underlying distribution. The learning step involves minimization of the generator's objective loss function,
\begin{equation}
\label{eq:loss_gen1}
\ell_{G}(\phi) = \mathbb E_{\boldsymbol x^\prime \sim G^{(i-1)}, \boldsymbol x\sim \mathcal D_{T^{(i)}}}\left[ L_{G} \left({\hat{\widetilde{\boldsymbol x} }}, \widetilde{\boldsymbol x}; \phi^{(i)}\right) \right],
\end{equation}
where ${\boldsymbol x^\prime \sim G^{(i-1)}}$ denotes the replay inputs drawn from the prior generator, ${\boldsymbol x\sim \mathcal D_{T^{(i)}}}$ denotes the input samples taken randomly from the dataset of the $i$-th task condition, ${L_{G}}$ is the loss function of the generator, and ${\hat{\widetilde{\boldsymbol x} }}$ is the reconstructed samples produced by the generator given the current setting of its parameter set, ${\phi^{(i)}}$. This step updates the generator of the $i$-th scholar.

\subsubsection{Tabular Variational Auto-Encoders}
TVAE is the state-of-the-art variational auto-encoder (VAE) for tabular data generation \cite{Xu2019ModelingTD}. It is specifically designed for tabular data, which is typically composed of a mix of both discrete and continuous features. Continuous features may have multi-modal and non-Gaussian distributions whereas discrete features are sometimes imbalanced making the modeling difficult. 

The basic scheme of a VAE composes of an encoder and a decoder \cite{Kingma2013AutoEncodingVB}. The encoder compresses the input ${\boldsymbol x}$ into the latent space. The decoder receives as input the information sampled from the latent space and produces ${\boldsymbol x^\prime}$ as similar as possible to ${\boldsymbol x}$. In TVAE, the modeling for encoder is similar to conventional VAE. The decoder is designed specially so that the probability distribution of data can be modeled accurately; refer to \cite{Xu2019ModelingTD} for details. Also, a mode-specific normalization is designed to deal with features with complicated distributions. This architecture allows TVAE to generate data that closely match the distribution of real tabular data, making it more suitable for applications like data imputation, synthetic data generation, and tabular data modeling.

\subsection{PROPOSED SOLUTION: MULTI-GENERATOR GENERATIVE REPLAY}
This section describes our construction of multi-generator generative replay. We first describe the construction of the generator selector and proceed with describing the construction of the multi-generator generative replay.

\subsubsection{Generator relevance determination}
\label{subsubsec:gen_rev}

Let ${\boldsymbol a= (a_1, \ldots, a_J)^\top}$, for all ${a_j\in\{0, 1\}}$, denote a binary vector of $J$ data configurations at a given task condition where each element specifies a different configuration, referred to as the \emph{task configuration vector}. 
In~\cite{rao2024generalizable}, Rao et al. showed that OWD distribution is significantly dependent upon the UE type. Therefore, in our use case of \ac{OWD} prediction,
the elements of the task configuration vector represent \ac{UE} device type; as an example, for three different \ac{UE} types as described in Table~\ref{tb:varied-param}, the task configuration vector is expressed by a 3-dimensional vector with binary elements with the three elements corresponding to the \ac{UE} types,
\begin{equation}
    \boldsymbol a= (\text{UE1}, \text{UE2}, \text{UE3})^\top.
\end{equation}
Further, we denote the task configuration vector at the $i$-th task condition by ${\boldsymbol a^{(i)}}$.

In a similar fashion, we define a configuration vector for the generators, named the \emph{generator configuration vector}. Formally, let ${\boldsymbol  b_k = (b_{k,1}, \ldots, b_{k, J})^\top}$, for all ${b_{k,j} \in[0,1]}$, denote the configuration vector for the $k$-th generator with the elements representing the same configurations as in the task configuration vector. However, unlike $\boldsymbol a$, elements of $\boldsymbol b$ do not need to be necessarily binary and can take soft values between zero and one; if the generator is fully applicable for a given configuration its corresponding element in $\boldsymbol b$ is set to one, if it is not applicable it is set to zero, and otherwise it may be set to a value in-between zero and one. 

We then define a vector of relevance scores for the $i$-th task condition, referred to as the \emph{relevance vector} and denoted by ${\boldsymbol r^{(i)}\in\mathbb R^K}$ where $K$ is equal to the number of generators. The elements of the relevance vector quantify how well a generator is relevant to the given task condition, and it is computed by the inner product of the task configuration vector and the generator configuration vector, expressed as:
\begin{equation}
\label{eq:rel_score}
\boldsymbol r^{(i)} = \left( {\boldsymbol b_1}^\top\boldsymbol a^{(i)}, \ldots, {\boldsymbol b_K}^\top\boldsymbol a^{(i)}\right)^\top.
\end{equation}
Finally, the index of the most relevant generator is given by:
\begin{equation}
\label{eq:gen_ind}
k_*  = \arg\!\max_{k} \left(\boldsymbol r^{(i)}\right), \quad \forall k=1,\ldots, K.
\end{equation}

In the context of our use case, from domain knowledge, we hypothesize that the UE device type is the major differentiating factor of data generation processes. As such for the first UE, the generator configuration vector is set to ${\boldsymbol b_1=(1,0,0)}$ which implies that this generator is suitable for the first UE device type. Similarly for the second and third UE devices, the generator configuration vectors are defined as ${\boldsymbol b_2=(0,1,0)}$ and ${\boldsymbol b_3=(0,0,1)}$, respectively.
As an example, let the task configuration vector at the current task be denoted by ${\boldsymbol a=({0, 1, 0)}}$ indicating the second UE device type. Then following \eqref{eq:rel_score} and \eqref{eq:gen_ind}, the generator defined by $\boldsymbol b_2$ would be the most relevant to the task condition defined by $\boldsymbol a$. In a similar fashion, upon arrival of a new task, the relevant generator will be selected based on the setting of its task configuration vector.

\begin{figure}[t!]
\centerline{\includegraphics[width=3.5in]{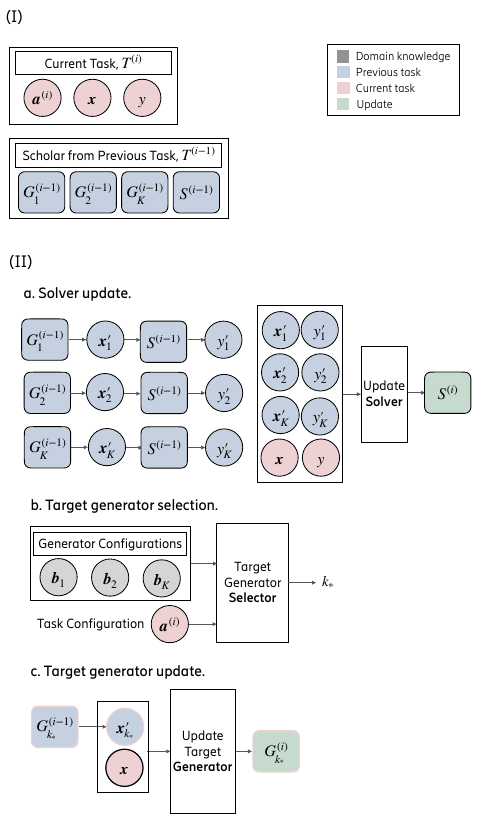}}
\caption{Multi-generator generative replay architecture. (I) Scholar from the previous task, and inputs and targets in the current task. (II) The procedure of learning of the scholar at the current task, which involves three steps.  
}\label{fig:method}
\end{figure}

\subsubsection{Construction of multi-generator generative replay}
We revisit the construction of the generative replay and extend it from a single generator to a finite set of multi-generators.

We define a scholar by a set of generators and a single solver, ${\mathcal H=\left\{\left\{G_k\right\}_{1:K}, S\right\}}$. Further, we assume their corresponding set of generator configuration vectors, $\left\{\boldsymbol b_k\right\}_{1:K}$, are given. The learning of the scholar at the current task condition, ${\mathcal H^{(i)}}$, involves a procedure of updating the solver followed by a generator selection step and a procedure of updating the generator. Fig.~\ref{fig:method} summarizes the proposed architecture for multi-generator generative replay with three steps procedure of learning of the scholar at the current task which is described in the following.

\paragraph{Solver update}
Let  ${\left\{\boldsymbol x^\prime_k\right\}_{1:K}}$ denote a set of replay inputs generated from ${\{G_k^{(i-1)}\}_{1:K}}$. A set of corresponding replay targets,  ${\left\{\boldsymbol y^\prime_k\right\}_{1:K}}$, are predicted from the prior state of the solver, ${S^{(i-1)}}$, where ${ y^\prime_k = S\left(\boldsymbol x^\prime_k; \theta^{(i-1)}\right)}$. Finally, given the input features from the current task condition, ${\boldsymbol x}$, the targets, $y$, and the replay targets, the solver's training loss is constructed according to:
\begin{multline}
\label{eq:loss_sol2}
\ell_{S} = \alpha \mathbb E_{(\boldsymbol x, y)\sim \mathcal D_{T^{(i)}}}\left[ L_S\left(S\left(\boldsymbol x; \theta^{(i)}\right), y \right)\right] \\ + (1-\alpha) \mathbb E_{\left\{\boldsymbol x_k^\prime\sim G_k^{(i-1)}\right\}_{1:K}}\left[ L_{S}\left( S\left(\boldsymbol x_k^\prime; \theta^{(i)} \right),  y_k^\prime \right)  \right].
\end{multline}

\paragraph{Target generator selection}
We need to obtain the index of the most relevant generator, ${k_*}$, for the current task configuration vector, ${\boldsymbol a^{(i)}}$, given $\left\{\boldsymbol b_k\right\}_{1:K}$. This is done by application of Eq.~\eqref{eq:rel_score} and Eq.~\eqref{eq:gen_ind}. The corresponding generator is referred to as the target generator at the $i$-th task condition and it is denoted by ${G^{(i-1)}_{k_*}}$.

\paragraph{Target generator update}
First, from the prior state of the target generator, ${G_{k_*}^{(i-1)}}$, replay input features are produced which are denoted by $\boldsymbol x_{k_*}^\prime$, where the subscript ${{k_{*}}}$ is used to emphasize on the generation through the target generator. The replay inputs are mixed with the input features from the current task condition, ${\boldsymbol x}$, and form ${\widetilde{\boldsymbol x}:=(\boldsymbol x, \boldsymbol x_{k_*}^\prime)}$. Given ${\widetilde{\boldsymbol x}}$, the target generator is updated via minimization of the following objective function:
\begin{equation}
\label{eq:loss_gen2}
\!\!\ell_{G_{k_*}}\!(\phi_{k_*}) \!= \!\mathbb E_{\boldsymbol x_{k_*}^\prime \!\sim G_{k_*}^{(i-1)}, \boldsymbol x\sim \mathcal D_{T^{(i)}}}\!\left[ L_{G_{k_*}} \!\!\left({\hat{\widetilde{\boldsymbol x} }}, \widetilde{\boldsymbol x}; \phi_{k_*}^{(i)}\right) \right],
\end{equation}
where ${\hat{\widetilde{\boldsymbol x} }}$ is the reconstructed samples produced by the target generator given the current setting of its parameter set, ${\phi_{k_*}^{(i)}}$.
This step updates the target generator of the $i$-th scholar while the other generators are unaltered.


\begin{figure*}
\centerline{\includegraphics[width=4.8in]{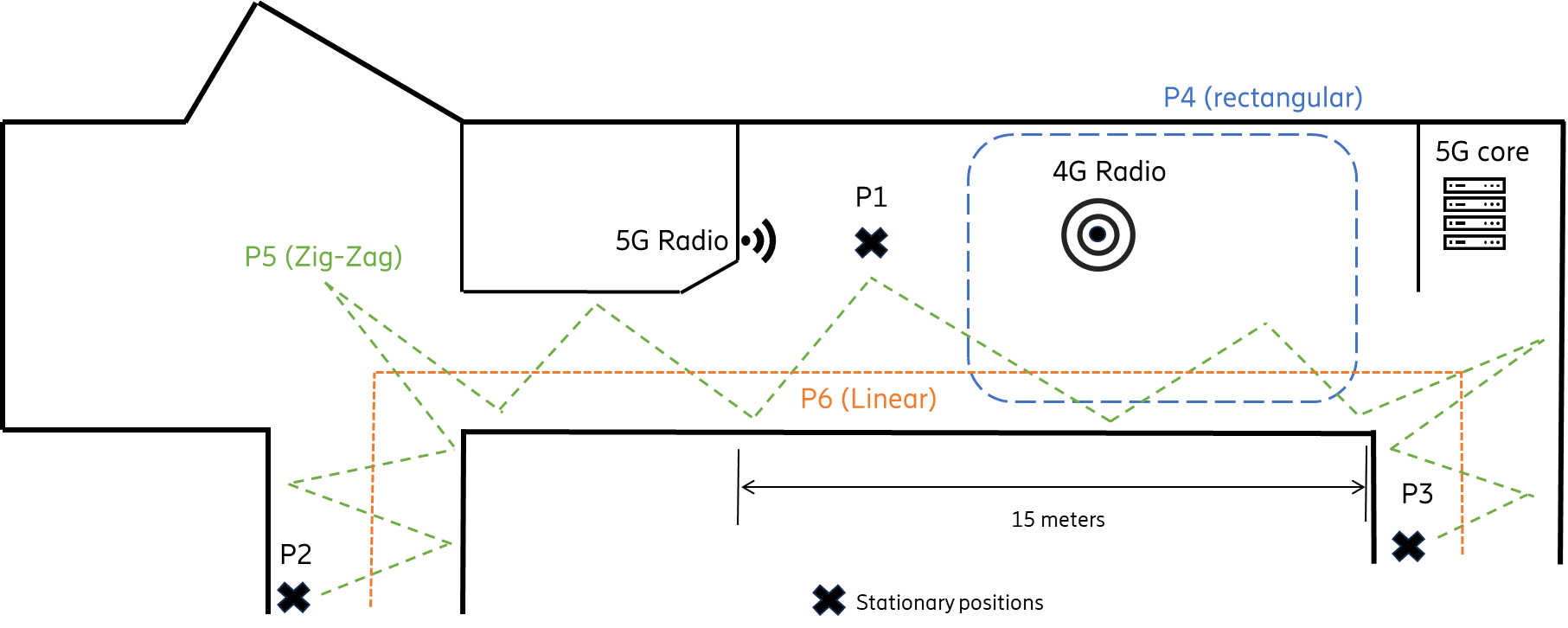}}
\caption{Floor plan of the testbed area illustrating positions and movement patterns (P1 - P6).}\label{fig:floorplan}
\end{figure*}

\section{A 5G TESTBED FOR CREATION OF ONE-WAY DELAY DATASETS}
\label{sec:dataset}

To evaluate the multi-generator generative replay approach for robust \ac{OWD} prediction in future 6G networks, we leverage an in-house 5G-mmWave testbed \cite{rao2024generalizable} to create datasets consisting of base station features and ground-truth \ac{OWD} values. In the following, while we briefly describe the testbed, we focus on the experiments conducted in support of evaluating our approach. 

The testbed is a 5G NSA system build upon commercially available Ericsson StreetMacro 6701 device that implements 5G NR, and an Ericsson Radio Dot 2243 device that implements 4G LTE. The StreetMacro does analog beamforming using one horizontal and one vertical beam (1 × 200 MHz dual polarized beams). We set the 4G LTE eNB to operate on band B3 (1800 MHz frequency, with 5 MHz bandwidth), and 5G NR \ac{gNB} on band n257 (28 GHz frequency, with 100 MHz bandwidth). 
The testbed resides in an indoor testbed area. The floor plan and associated positions of the 5G NR and 4G LTE radios, is illustrated in Fig.~\ref{fig:floorplan}. For details regarding the setup we refer to \cite{rao2024generalizable}. 

Similarly to \cite{rao2024generalizable}, we measure \ac{OWD}, between the \ac{UE} and a receiver in the network, with three different types of 5G \ac{UE} devices, with different manufacturers and chipsets, referred to as \ac{UE}1, \ac{UE}2, \ac{UE}3. The \ac{OWD} measurements are conducted using TWAMP \cite{hedayat2008two} probe packets of sizes 100 and 1400 bytes. An additional \ac{UE} is configured for creation of network traffic scenarios, where that \ac{UE} is introducing competing uplink traffic into the network, ranging from 0 to 60 Mbps, in steps of 10 Mbps. A \ac{UE} is either stationary (position 1, 2, or 3), or moving according to a set of predefined movement patterns (rectangular, zigzag, and linear). The experiment parameters are summarized in Table \ref{tb:varied-param}.  

The experimental scenarios were designed to expose the \ac{UE} and the network to a wide range of conditions that contribute to variations in \ac{OWD} and baseband features. The range of loads was selected to expose the network from being completely empty, to operating in a state of congestion. In addition, the \ac{UE} movement patterns were designed to ensure that we captured excellent channel conditions, worst-case cell edge conditions, as well as conditions in between. These network conditions, when put into a sequence, represent distributional shifts that the multi-generator generative replay approach must mitigate.  

In addition to collection of ground-truth \ac{OWD}, the testbed was configured with multiple measurement points to extract network metrics \cite{rao2024generalizable}. More specifically, we extract logs from the \ac{gNB} corresponding to beam forming, \ac{UE} connectivity, and \ac{UL}/\ac{DL} scheduling events~\cite{3gpp.38.214,3gpp.38.215,3gpp.38.321}, on the sub-millisecond scale. Of a much larger set available, we choose to monitor and use 103 metrics.  

For each set of parameters listed in Table \ref{tb:varied-param}, we sent 5000 TWAMP packets, spacing each packet 50 ms apart. This process generated a dataset with 1.26 million accurate \ac{OWD} samples. We averaged the network metrics obtained from the \ac{gNB} over 50 ms, and organized them into bins aligned in time with the \ac{OWD} values. Consequently, each sample in the dataset includes a \ac{OWD} measurement and the average value of each feature within the 50 ms bin. 

In this paper, we focus on \ac{OWD} prediction for \ac{UL}, while the approaches can be generalized to DL as well. Relevant network metrics are used as features. We use measurement experiments data with all varying \ac{UL} load and large probe packet size (1400B).

\begin{table}
\scriptsize
    \centering
    \caption{Testbed experiment parameters. }
    \begin{tabular}{|p{2.2cm}|p{4.9cm}|}
        \hline
        \textbf{\ac{UE} device type} & \ac{UE}~1, \ac{UE}~2, \ac{UE}~3 \\
        \hline
        \textbf{Probe pkt. size} & \SI{100}{\byte}, \SI{1400}{\byte}\\
        \hline
        \textbf{UL load} & \SIlist{0;10;20;30;40;50;60}{\Mbps} \\ 
        \hline
        \multirow{4}{*}{\parbox[t]{5cm}{\textbf{Position and}\\\textbf{ Movement patterns}}} 
        & P1: Stationary at positions 1\\
        \cline{2-2}
        & P2: Stationary at positions 2\\
        \cline{2-2}
        & P3: Stationary at positions 3\\
        \cline{2-2}
        & P4: Rectangular movement\\ 
        \cline{2-2}
        & P5: Zigzag movement, pos. $1\rightarrow 2\rightarrow 3\rightarrow 1$\\
        \cline{2-2}
        & P6: Linear movement, pos. $1 \rightarrow 2 \rightarrow 3 \rightarrow 1$\\
        \hline
        \end{tabular}
        \vspace{-1em}
        \label{tb:varied-param}
    \end{table}

\section{EVALUATION METHODOLOGY}
\label{sec:evaluation}
In this section, we describe our evaluation methodology, starting with describing our approach in designing CL task sequences, presenting methods considered in our evaluation framework, and introducing relevant metrics devised for assessing the performance of the CL methods.

\subsection{DESIGNING CL TASK SEQUENCES}
\label{subsec:task_seq}

In~\cite{rao2024generalizable}, Rao et al. showed that various configuration settings such as \ac{UE} device type and  the \ac{UE} movement pattern can result in distributional shift both at the input space as well as in the task space (\ac{OWD}). Inspired by this, we create a series of CL tasks by utilizing configuration settings from various experiments to define the tasks. In particular, each task is constructed based on the UE type, along with its position and movement pattern as defined and illustrated in Table~\ref{tb:varied-param} and Fig.~\ref{fig:floorplan}. Accordingly, two groups of CL task sequences are constructed which are shown in Table~\ref{tb:cl-seq}.

In the first group, named Group 1, we assume there are only two different UE device types along with 6 different positions and movement patterns. Table~\ref{tb:cl-seq} (a) summarizes 6 different scenarios considered in our evaluations each labeled by a case identifier (Case ID). The scenarios are devised based on the UE device types and their movement patterns. As an example, in one scenario (Case ID 1), UE 1 and UE 2 are considered together with the UE movement pattern from stationary to moving ({$S \rightarrow M$}). To explore the impact of task order on the evaluation results, in the other scenario (Case ID 2), the order of task sequence of UE 1 and UE 2 is reversed from moving to stationary ({$M \rightarrow S$}). 
In the same way, the task sequence of UE 2 and UE 3 (Case IDs 3,4) and UE 3 and UE 1 (Case IDs 5,6) are designed. All scenarios in Group 1 consists of 12 CL tasks in each. Among the various order of tasks we explored, the ones presented here are representative of the overall patterns observed.

Similarly, in the second group, named Group 2, we assume there are three different UE device types with 6 different positions and movement patterns. Two scenarios (Case IDs 7,8) that are considered in the evaluations are summarized in Table~\ref{tb:cl-seq} (b). Both scenarios in Group 2 consists of 18 CL tasks in each.

\begin{table*}[]
\scriptsize
\caption{CL task sequences for Group 1 and Group 2. Each task sequence is labeled by Case ID and Case Name. CL tasks are devised based on the UE device types, along with its position and movement pattern. Note, P is for position and movement pattern, whereas $S$ is for stationary, and $M$ for moving.
}.
\centering
\begin{tabular}{lllllllllllllll}
(a) Group 1\\
\hline
Case ID            & Case Name                   & CL task  & 1 & 2 & 3 & 4 & 5 & 6 & 7 & 8 & 9 & 10 & 11 & 12 \\ \hline\hline
\multirow{2}{*}{1} & UE 1,2                        & UE       & 2 & 1 & 2 & 1 & 2 & 1 & 2 & 1 & 2 & 1 & 2 & 1\\ 
                   & {$S \rightarrow M$}                     & P & 1 & 1 & 3 & 3 & 2 & 2 & 4 & 4 & 6 & 6 & 5 & 5\\ 
\hline 
\multirow{2}{*}{2} & UE 1,2                        & UE       & 1 & 2 & 1 & 2 & 1 & 2 & 1 & 2 & 1 & 2 & 1 & 2\\ 
                   & {$M \rightarrow S$}                     & P & 5 & 5 & 6 & 6 & 4 & 4 & 2 & 2 & 3 & 3 & 1 & 1\\ 
\hline
\multirow{2}{*}{3} & UE 2,3                        & UE       & 2 & 3 & 2 & 3 & 2 & 3 & 2 & 3 & 2 & 3 & 2 & 3\\ 
                   & {$S \rightarrow M$}                     & P & 1 & 1 & 3 & 3 & 2 & 2 & 4 & 4 & 6 & 6 & 5 & 5\\ 
\hline
\multirow{2}{*}{4} & UE 2,3                        & UE       & 3 & 2 & 3 & 2 & 3 & 2 & 3 & 2 & 3 & 2 & 3 & 2\\ 
                   & {$M \rightarrow S$}                     & P & 5 & 5 & 6 & 6 & 4 & 4 & 2 & 2 & 3 & 3 & 1 & 1\\ 
\hline
\multirow{2}{*}{5} & UE 1,3                        & UE       & 3 & 1 & 3 & 1 & 3 & 1 & 3 & 1 & 3 & 1 & 3 & 1\\ 
                   & {$S \rightarrow M$}                     & P & 1 & 1 & 3 & 3 & 2 & 2 & 4 & 4 & 6 & 6 & 5 & 5\\ 
\hline
\multirow{2}{*}{6} & UE 1,3                        & UE       & 1 & 3 & 1 & 3 & 1 & 3 & 1 & 3 & 1 & 3 & 1 & 3\\ 
                   & {$M \rightarrow S$}                     & P & 5 & 5 & 6 & 6 & 4 & 4 & 2 & 2 & 3 & 3 & 1 & 1\\ \hline 
\end{tabular}
\\
\vspace{2ex}
\begin{tabular}{lllllllllllllllllllll}
(b) Group 2
\\
\hline
Case ID            & Case Name                      & CL task  & 1 & 2 & 3 & 4 & 5 & 6 & 7 & 8 & 9 & 10 & 11 & 12 & 13 & 14 & 15 & 16 & 17 & 18\\ \hline\hline
\multirow{2}{*}{7} & UE 1,2,3                          & UE       & 2 & 3 & 1 & 2 & 3 & 1 & 2 & 3 & 1 & 2 & 3 & 1 & 2 & 3 & 1 & 2 & 3 & 1\\ 
                   & {$S \rightarrow M$}                        & P & 1 & 1 & 1 & 3 & 3 & 3 & 2 & 2 & 2 & 4 & 4 & 4 & 6 & 6 & 6 & 5 & 5 & 5\\ 
\hline
\multirow{2}{*}{8} & UE 1,2,3                          & UE       & 1 & 3 & 2 & 1 & 3 & 2 & 1 & 3 & 2 & 1 & 3 & 2 & 1 & 3 & 2 & 1 & 3 & 2\\ 
                   & {$M \rightarrow S$}                        & P & 5 & 5 & 5 & 6 & 6 & 6 & 4 & 4 & 4 & 2 & 2 & 2 & 3 & 3 & 3 & 1 & 1 & 1\\ \hline
\end{tabular}
\label{tb:cl-seq}
\end{table*}

\begin{table*}[]
\caption{Neural network architecture used in evaluations.}
\scriptsize
\centering
\begin{tabular}{l|l}
\hline
\multirow{2}{*}{Generator VAE}   & Encoder: Hidden layers sizes: (128, 128). Activation function: ReLU. \\                          
                                 & Decoder: Hidden layers sizes: (128, 128). Activation function: ReLU. \\
\hline
\multirow{2}{*}{Generator TVAE}  & The TVAE Synthesizer in python library SDV \cite{SDV} is used to train a TVAE model and generate synthetic data. \\
                                 & Default parameters are applied, e.g. size of each hidden layer in the encoder and decoder is (128, 128).\\
\hline
\multirow{2}{*}{Solver MLP}      & Hidden layers sizes: (200, 150, 100, 50). Activation function: ReLU. \\
                                 & Loss function: MSE loss. Optimizer: Adam. Learning rate: 0.001. \\

\hline
\end{tabular}
\label{tb:nn-architect}
\end{table*}

\subsection{EVALUATION FRAMEWORK}
\label{subsec:eva_fram}

We evaluate our multi-generator generative replay approach using the two defined groups of CL task sequences, and compare to several baselines as described below. 
For approaches based on generative replay, we employ a fully connected MLP neural network as the solver.
The architecture of the MLP is shown in Table~\ref{tb:nn-architect}. In the following, we discuss 4 baseline approaches, as well as providing additional details regarding our own approach. 

\subsubsection{Baseline approaches}
\paragraph{Naïve}
As in \cite{Maltoni2018ContinuousLI}, Naïve method could be used as the lower bound baseline for stability evaluation, in which the model is trained continuously with data of each task, without any particular framework to control forgetting. That is, the solver is finetuned using data from the new task, without access to the data from the previous tasks.

\paragraph{Cumulative}
Cumulative method is used to evaluate the upper bound of forgetting in CL as in \cite{Maltoni2018ContinuousLI}. For each task, it accumulates all data from previous tasks and data of current task, re-train the solver from scratch. There is no forgetting as cumulative method could access and use all the previous data. However, it is often infeasible in real world applications, due to constrains in data storage and privacy.

\paragraph{Single generator with VAE (SingleGen-VAE)} 
It is illustrated in the state-of-the-art deep generative replay algorithm\cite{Shin2017ContinualLW} that a VAE could be used as the generator. We evaluate its performance on our cases. The architecture of the VAE is shown in Table~\ref{tb:nn-architect}. 
It is worth noting that we experimented with different numbers of layers and layer sizes where all of which produced similar results.

\paragraph{Single generator with TVAE (SingleGen-TVAE)} 
As it is presented in Section~\ref{sec:approach}, TVAE is specifically designed for tabular data generation. The dataset of this study is created from the 5G testbed, and the features of relevant network metrics are a mix of both discrete and continuous columns. 
Considering the tabular nature of the data, we view TVAE as a more appropriate choice compared to VAE. This aligns with one of the objectives of this work, which is to highlight the significance of selecting the right generative model based on the data structure. 
The architecture of the TVAE is shown in Table~\ref{tb:nn-architect}. 

\subsubsection{Multi-generator with TVAE (MultiGen-TVAE)} 
As it is illustrated in Section~\ref{sec:approach}, based on domain knowledge, UE device type is the primary determinant of how data is generated. The generator configuration vectors ${\boldsymbol b_1=(1,0,0)}$,${\boldsymbol b_2=(0,1,0)}$ and ${\boldsymbol b_3=(0,0,1)}$ are defined for three UE device types as described in Section~\ref{sec:approach}. For CL task sequence Group 1, there are only two different UE device types in each scenario, which leads to two generators, one for odd-numbered tasks and the other for even-numbered tasks. For CL task sequence Group 2, there are three different UE device types in each scenario, which leads to three generators.
Here we consider TVAE as our choice of the generative model.

\subsection{PERFORMANCE EVALUATION}

As the prediction of OWD is a regression task, the performance of the model is evaluated using a standard percentage error metric known as mean absolute percentage error (MAPE), aligning with our prior study~\cite{rao2024generalizable}, thereby facilitating consistent benchmarking and comparative analysis. MAPE is defined as:
\begin{equation}
\mathrm{MAPE} = \frac{100}{N}\sum_{i=0}^{N-1}\frac{\left | y_i - \widehat{y}_i\right |}{\left | y_i \right | },
\end{equation}
where $y_i$ is the ground truth, $\hat{y_i}$ is the predicted \ac{OWD}, and $N$ is the number of samples in the task dataset. 
Lower MAPE values indicate better model performance.

To evaluate the performance of the selected CL algorithms, we split the dataset for each CL task into training and test sets (70\% for training). Training is done sequentially with the training set of each task according to the task sequence. Performance is then evaluated using the test sets for all $I$ tasks.
After the model finishes learning of all $I$ tasks, we get the result matrix ${R \in \mathbb R^{I,I}}$, where $R_{i,j}$ is the performance of the model on test set of task $T_{j}$ after training with the training set of task $T_{i}$. Specifically for our evaluations, the model performance $R_{i,j}$ is MAPE values. Inspired by the two CL evaluation metrics Average Accuracy (ACC) and Backward Transfer (BWT) in \cite{ LopezPaz2017GradientEM}, we define metrics Average MAPE (AveMAPE) and Forgetting(F) for our regression task OWD prediction, where

\begin{equation}
\mathrm{AveMAPE} = \frac{1}{I}\sum_{i=1}^{I}R_{I,i},
\end{equation}
\begin{equation}
\mathrm{F} = \frac{1}{I-1}\sum_{i=1}^{I-1}R_{I,i}-R_{i,i}.
\end{equation}
AveMAPE evaluates the overall performance of the tasks learned so far. F shows the influence of learning the current task on the performance of the previous tasks. In other words, it evaluates the memory stability of old tasks. 
Higher values of F indicate increased forgetting.

Furthermore, here, we introduce a new metric to assess effect of both short-term and long-term forgetting, since F alone does not capture short-term and long-term forgetting specifically. We refer to this metric as ${\mathrm{F}_k}$, which effectively measures average k-step forgetting and is defined as:

\begin{equation}
\mathrm{F}_k = \frac{1}{I-k}\sum_{i=1}^{I-k}R_{i+k,i}-R_{i,i}, \forall k=1, \ldots, I-1,
\end{equation}
where for smaller values of $k$, ${\mathrm{F}_k}$ reflects short-term forgetting while for larger values of $k$, it captures the long-term forgetting.


\section{RESULTS}
\label{sec:results}

\begin{figure*}[t!]
\centering
\includegraphics[width=0.75\linewidth]{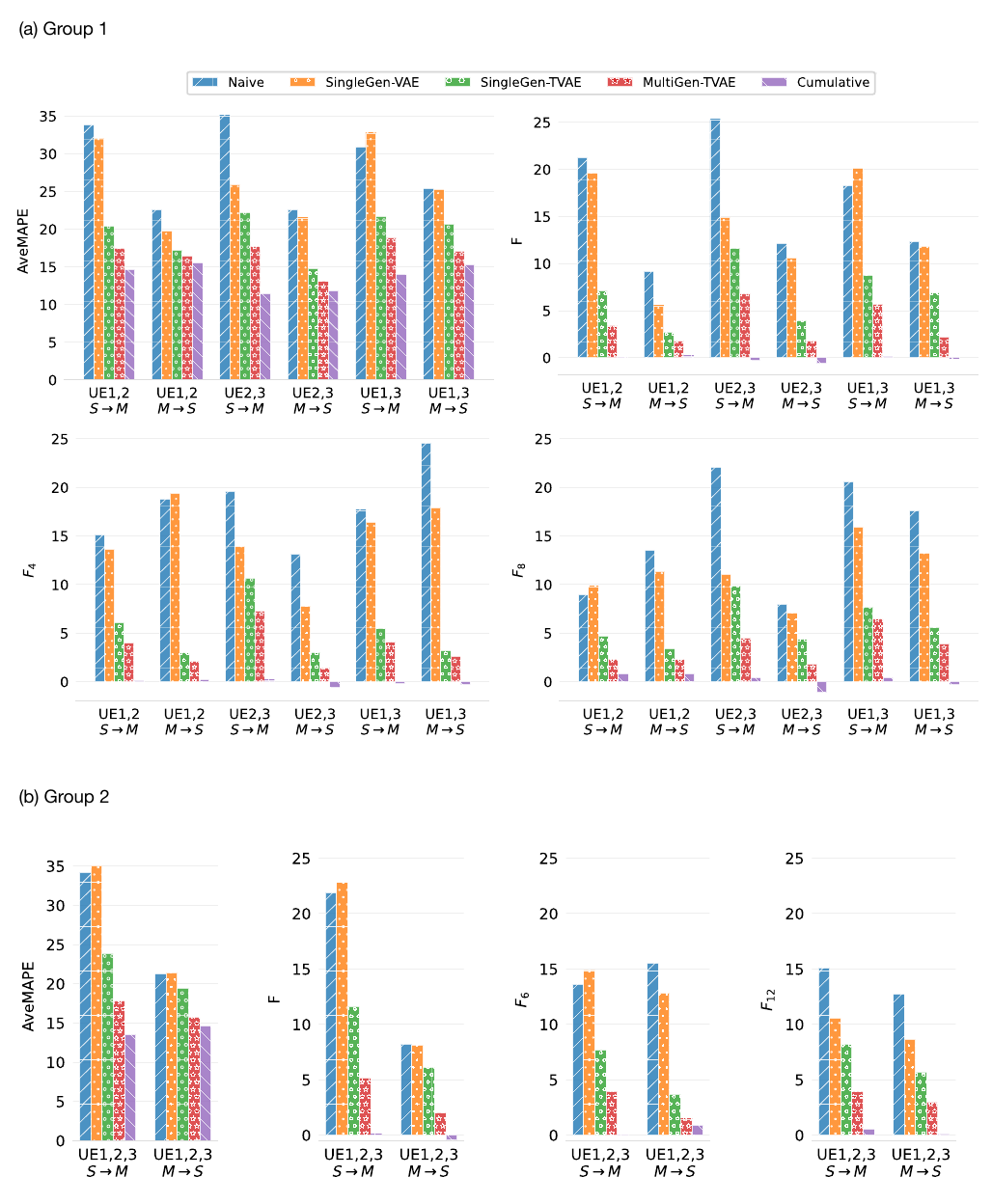}
\caption{Performance evaluation on CL task sequence Group 1 and Group 2.}
\label{fig:12_18_tasks_bar}
\end{figure*}

\begin{figure*}[t!]
\centering
\includegraphics[width=0.7\linewidth]{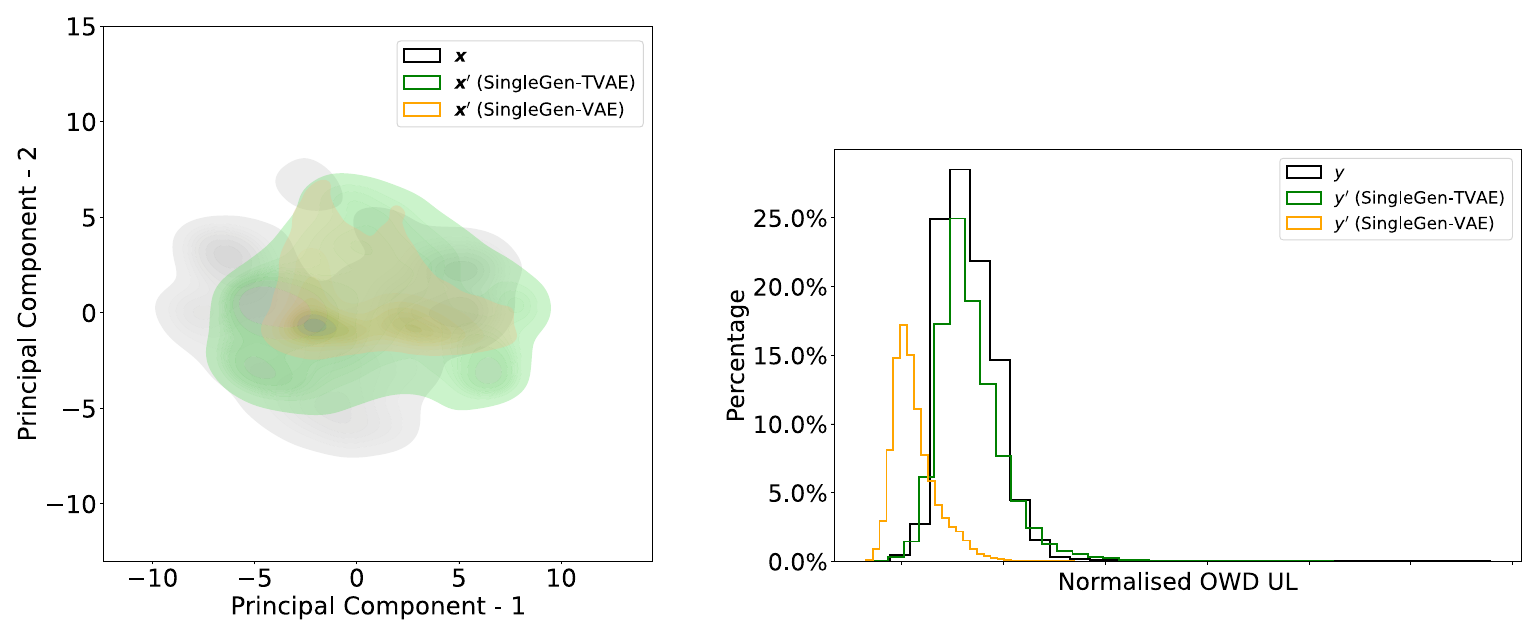}
\caption{KDE plot of principal components on generated ${\boldsymbol x^\prime}$ by TVAE and VAE and histogram on corresponding ${y^\prime}$.}
\label{fig:pca_hist_tvae}
\end{figure*}

\begin{figure*}[t!]
\centering
\includegraphics[width=0.8\linewidth]{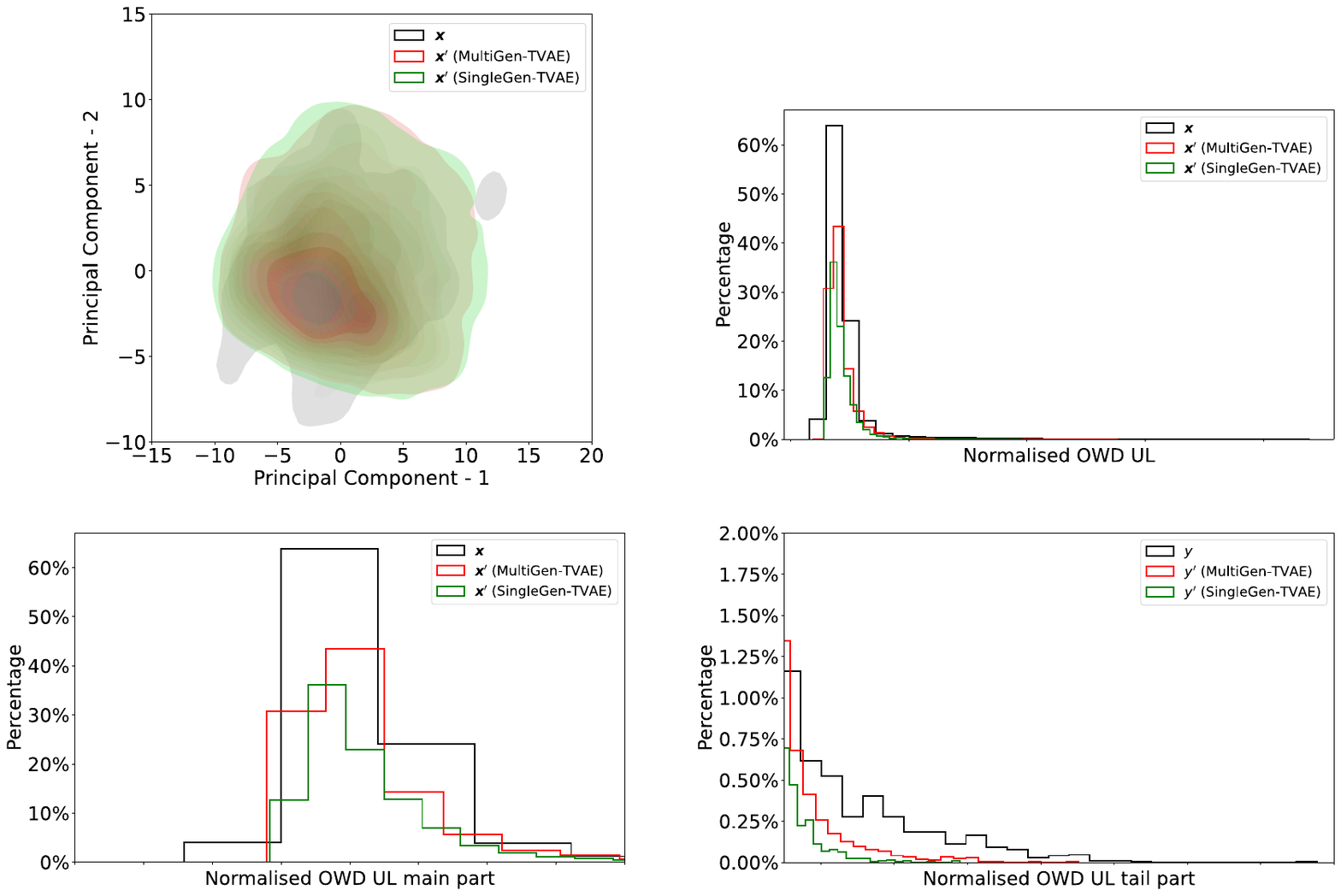}
\caption{KDE plot of principal components on generated ${\boldsymbol x^\prime}$ by multi-generator and single generator and histogram on corresponding ${y^\prime}$, and zoom in on main part and tail part. }
\label{fig:pca_hist_multiG}
\end{figure*}

\begin{figure*}[t!]
\centering
\includegraphics[width=0.85\linewidth]{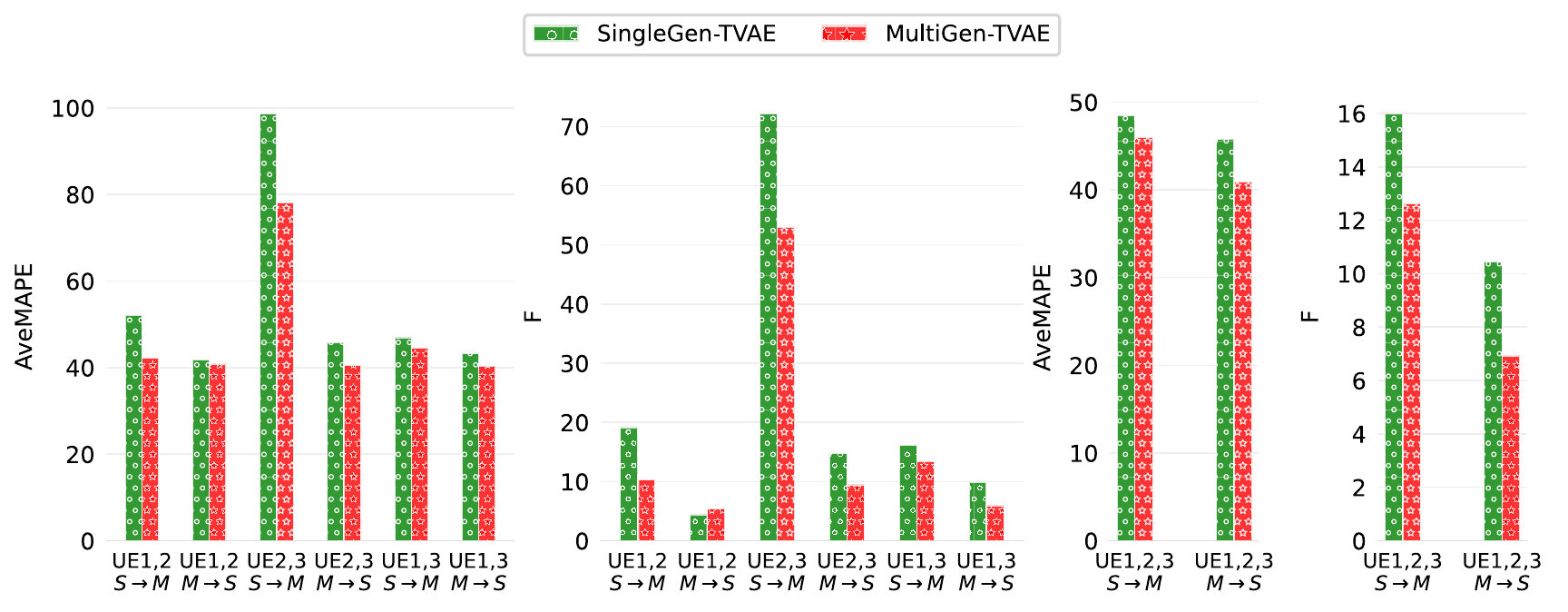}
\caption{Performance evaluation on \ac{OWD} tail in CL task sequence Group 1 and Group 2 by metrics AveMAPE, F.}
\label{fig:tail_bar_plot}
\end{figure*}

\begin{figure}[t!]
\centering
\includegraphics[width=0.55\linewidth]{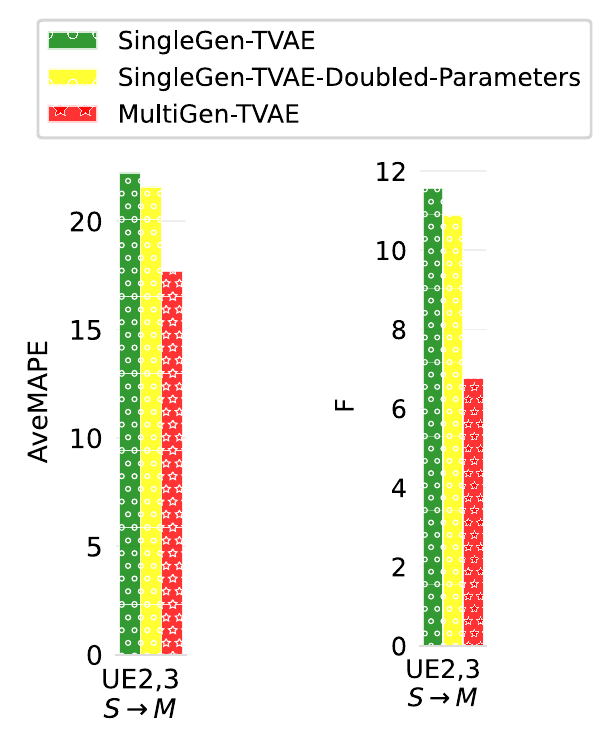}
\caption{Performance evaluation on \ac{OWD} by metrics AveMAPE, F with doubled size of generator parameters.}
\label{fig:doubled_parameter_bar_plot}
\end{figure}

In this section, we first present the evaluation results of the CL task sequences Group 1 and Group 2, defined  in Section~\ref{sec:evaluation}.
To further analyze our observations, in the following, we evaluate the distribution of generated ${\boldsymbol x^\prime}$ and corresponding ${y^\prime}$ by VAE and TVAE, single generator and multi-generator.
In addition, we evaluate the model performance on the tail of the OWD distribution. Finally, the model performance of single generator and multi-generator with the equal number of learnable parameters are compared.

\subsection{EVALUATION ACROSS GROUP 1 AND GROUP 2}
Fig.~\ref{fig:12_18_tasks_bar} summarizes the model performance across methods, our multi-generator approach as well as the baselines. The performance of the model is evaluated by the metrics AveMAPE, F, and selected F$_k$ reflecting of short-term and long-term forgetting: namely, F$_4$ and F$_8$ for Group 1 with 12 tasks in each sequence, and F$_6$ and F$_{12}$ for Group 2 with 18 tasks in each sequence.

The results notably demonstrate that our proposed multi-generator based approach, MultiGen-TVAE, consistently outperforms the single-generator based approaches, SingleGen-TVAE and SingleGen-VAE, across all metrics and scenarios. Next, we observe that the choice of generative model considerably impacts overall performance; notably, the TVAE-based approach SingleGen-TVAE clearly outperforms the alternative VAE-based approach SingleGen-VAE. This suggests that VAE struggles to generate sufficiently representative data samples for previous tasks, potentially introducing noise with the generated samples. Additionally, as expected, the figures show that the Naïve method generally represents the lower bound of performance across various scenarios, whereas the Cumulative method consistently reaches the upper bound of performance in all scenarios, consistent with the discussion in Section~\ref{sec:evaluation}. It is also apparent that the order of tasks affects the results; different sequences of tasks lead to variations in overall performance and forgetting in CL.

\subsection{VAE VS TVAE}

During the model training phase for Task 1, the generator is trained using ${\boldsymbol x}$, while the solver (OWD model) is trained with $({\boldsymbol x, y})$. To compare VAE and TVAE, we generate ${\boldsymbol x^\prime}$ for Task 1 using both VAE and TVAE. As shown in Fig.~\ref{fig:pca_hist_tvae}, we then apply principal component analysis (PCA) with 2 principal components on the generated ${\boldsymbol x^\prime}$ and visualize these components using a kernel density estimate (KDE) plot. Additionally, we plot the histogram of the corresponding ${y^\prime}$ by applying the OWD model to the generated ${\boldsymbol x^\prime}$. We take the CL task sequence with Case ID 1 as an example.  As demonstrated in Fig.~\ref{fig:pca_hist_tvae}, the distribution of ${\boldsymbol x^\prime}$ and ${y^\prime}$ generated by TVAE closely resembles the real (${\boldsymbol x}, y$) of Task 1, compared to that generated by VAE. This indicates that TVAE is more effective in generating data samples that better simulate real samples, which allows SingleGen-TVAE to provide superior data for replay during CL model training, resulting in improved performance in OWD prediction compared to SingleGen-VAE.

\subsection{SINGLE GENERATOR VS MULTI-GENERATOR}
\label{subsec:single_multi}
In the model training phase before training Task 12, the generator or generators is/are trained with ${\boldsymbol x}$ of Task 1 to Task 11, and the solver (OWD model) is trained with $({\boldsymbol x, y})$ of Task 1 to Task 11. To compare single generator and multi-generator, we generate ${\boldsymbol x^\prime}$ of Task 1 to Task 11 by single generator and multi-generator. As shown in Fig.~\ref{fig:pca_hist_multiG}, we then perform PCA with 2 principal components on generated ${\boldsymbol x^\prime}$ and visualize these components using the KDE plot. Additionally, we plot the histogram of the corresponding ${y^\prime}$ by applying \ac{OWD} model on generated ${\boldsymbol x^\prime}$. We take the CL task sequence with Case ID 1 as an example. 
As demonstrated in Fig.~\ref{fig:pca_hist_multiG}, with a focus on the main and tail parts of the histogram of ${y^\prime}$, the data distribution of ${y^\prime}$ generated by the multi-generator more closely resembles the real ${y}$ compared to that generated by the single generator. As multi-generator could generate data samples that simulate the real samples better, MultiGen-TVAE has better generated data samples to replay for training of CL model, resulting in enhanced performance in OWD prediction compared to SingleGen-TVAE.

\subsection{EVALUATION ON OWD TAIL}
In addition to assessing OWD prediction across all data samples, it is important to evaluate the tail of the OWD distribution. To do this, we set the threshold and select the data samples with \ac{OWD} values larger than it, then calculate the evaluation metrics using only these selected samples. Fig.~\ref{fig:tail_bar_plot} presents the evaluation results for the tail of the OWD distribution using metrics AveMAPE and F with CL task sequences Group 1 and Group 2. The results indicate that, for nearly all task sequences, the proposed MultiGen-TVAE method outperforms SingleGen-TVAE in terms of tail performance for both AveMAPE and F metrics. This observation aligns with the findings in Section~\ref{sec:results} C that multi-generator could generate data samples that simulate the real samples better compared with single generator for OWD tail part.

\subsection{COMPARISON OF GENERATORS WITH SAME SIZE}
To compare the performance of SingleGen-TVAE and MultiGen-TVAE with generative models having an equivalent number of learnable parameters, we doubled the parameters for SingleGen-TVAE, as outlined in Table~\ref{tb:nn-architect}. Specifically, we increased each hidden layer in SingleGen-TVAE from 128 to 256 units. The evaluation was conducted using the CL sequence, with Case ID 3 as an example. As illustrated in Fig.~\ref{fig:doubled_parameter_bar_plot}, MultiGen-TVAE outperforms SingleGen-TVAE, despite the doubling of generator parameters in SingleGen-TVAE.


\section{DISCUSSION}
\label{sec:discussion}

\label{subsec:approach_diss}

From a methodological perspective, in the context of \ac{OWD} prediction, we have made several key contributions aimed at mitigating the risk of catastrophic forgetting in CL. Firstly, we emphasized the significance of utilizing a tailored generative model to ensure that the data synthesis process aligns with the structural characteristics and specific requirements of the data. Given our use case involving tabular data, we adopted a generative model specifically designed for this type of data, the TVAE. This choice was shown to significantly enhance the quality of generated samples needed for effective replay and the retention of learned information.
Next, we designed a multi-generator architecture for generative replay to tackle the challenges posed by single-generator systems, particularly in the presence of concept drift within the CL framework. By employing multiple generators, it was shown that we can reduce the effects of concept drift since each generator would be able to concentrate on producing samples tailored to a distinct concept, thereby enhancing the preservation of task-specific knowledge.
Finally, to address the challenge of generator selection within our multi-generator generative replay framework, we implemented a domain-guided selection mechanism that utilizes domain knowledge for optimal task assignment to generators. In our use case, the differentiating factor for generator selection was the UE device type, provided as domain knowledge. This is one of the potential task configuration options and that other possibilities could be studied in future works. Moreover, moving forward, exploring data-driven approaches for generator selection presents a promising direction for future research. Through comprehensive empirical experiments, we demonstrated that the steps above significantly reduce risks of catastrophic forgetting and enhance the model's capability to sustain performance across a diverse array of tasks. By maintaining high-quality data replay and strategic task allocation, our approach enhances adaptability in dynamic environments.

Replay has been shown to be an effective strategy in CL if performance is the main objective. However, it requires large memory and often infeasible in real world applications where the access to past data is limited due to privacy-preserving. Instead of saving raw data, generative replay is a competitive alternative approach. As an example, the size of the raw data of task sequence with Case ID 1 is 170 MB. The size of one TVAE generator is 2.4 MB, and multi-generator with two TVAE generator in this case is 4.8 MB. When the number of tasks increase, the size of raw data increases linearly, while the size of generator won't change.
This illustrates the benefits of our approach from an energy perspective. Our results show that the current single generator approach does not work well, in order to make it comparable to the multi-generator approach taken here it is reasonable to expect that it would get very costly from an energy perspective as this would require more training data and model parameters in the generator. For future work, it would be interesting to look closer at the generator training; how much energy does the training of the generator take compared to Replay-Based approaches with sample selection, and how large in terms of data, parameters and training would a successful single generator approach be. Furthermore, studying the energy consumption of this method compared to simple baselines such as isolated training for each task would be interesting.


\section{RELATED WORK}
\label{sec:relatedwork}
CL methods have been proposed to improve various aspects of ML and has been categorized in different ways
\cite{Wang2023ACS}\cite{lesort2019continuallearningroboticsdefinition}\cite{DeLange2019ACL}. In \cite{Wang2023ACS}, based on how task specific information is stored and used throughout the sequential learning process, the authors distinguish five major categories, which are Regularization-Based approach, Replay-Based approach, Optimization-Based approach, Representation-Based approach and Architecture-Based approach.
The different categories come with different properties related to energy usage as described in \cite{trinci2024green}, where the authors compare different CL approaches in terms of energy consumption. Here the Representation-Based approaches score highest on energy efficiency but other approaches such as Replay-Based still outperform the naive baseline of joint training.

The focus of our paper is generative replay or pseudo-rehearsal, which is a sub-direction in the Replay-Based approach. Instead of storing old training samples, generative replay requires training a generative model to replay generated data. Compared with other approaches, one benefit of generative replay is that the generative model makes it possible to provide data samples from previous tasks for future needs, with much less memory and feasible for privacy concerns. 
As it is presented in Section \ref{sec:approach}, DGR\cite{Shin2017ContinualLW} establishes a framework that the learning of each new task is coupled with replaying generated data sampled from one generative model, ensuring the retention of previously acquired knowledge. MeRGAN \cite{Wu2018MemoryRG} incorporates a memory replay mechanism to prevent catastrophic forgetting, which is further enforces through either joint training or replay alignment.

CL approaches in various categories have advantages and work best within specific scenarios. Between the categories, approaches are often orthogonal with respect to each other. Hence, generative replay could be hybrid with other CL strategies. To mitigate catastrophic forgetting of generative models, VCL\cite{Nguyen2017VariationalCL} uses a weight regularization based variational Bayesian approach, and maintaining a distribution over parameters and updating it using evidence lower bound (ELBO). DGM \cite{Ostapenko2019LearningTR} is inspired by biological synaptic plasticity principles, incorporating mechanisms like memory consolidation and task-specific adaptations.

For pseudo-rehearsal, the generative models could be of various types, such as generative adversarial networks (GANs) and variational autoencoder (VAE). GANs are typically used in scenarios where high-quality, realistic samples are required, such as image generation, super-resolution, and artificial data generation \cite{Ostapenko2019LearningTR}. GANs are highly flexible and can be adapted to various types of generative tasks, but can be harder to control due to the adversarial nature. VAEs are more suited for tasks where a structured latent space and stable training are important \cite{Kemker2017FearNetBM} \cite{Riemer2017ScalableRF}. Since VAEs have an explicit, interpretable latent space, they are often more useful in applications where understanding the latent factors of data is important. Besides GANs and VAE, the novel approach DDGR \cite{Gao2023DDGRCL} adopts a diffusion model as the generator and calculates an instruction-operator through the classifier to instruct the generation of samples. \\


\section{CONCLUSION}
\label{sec:conclusions}
In this paper, we present a novel approach which introduces the concept of multi-generator for the state-of-the-art CL generative replay framework, along with TVAE as generative models. We emphasized the significance of utilizing a tailored generative model and producing samples tailored to a distinct concept by multi-generator, to enhance the preservation of data and task-specific knowledge. For our use case, the domain knowledge of \ac{UE} capabilities is incorporated into the learning process for determining generator setup and relevance. The proposed approach is evaluated across a diverse set of scenarios with data that is collected in a realistic 5G testbed, demonstrates mitigation of catastrophic forgetting of \ac{OWD} prediction and tail prediction model, in comparison to baselines. Strategic task allocation combined with high-quality data replay empowers our approach to adapt more effectively in dynamic environments. Furthermore, this approach reduces the need for data storage efficiently, addressing the challenges of resource constraints in 5G networks. 

For future research, one promising direction is to explore data-driven approaches for generator selection, and to evaluate the performance in comparison to domain-guided selection.  In addition, future work may benefit from investigate alternative generative model architectures, such as temporal generative models, especially when OWD datasets are represented as time series data.


\section*{ACKNOWLEDGMENT}
This research was supported by the Swedish Governmental Agency for Innovation Systems (VINNOVA) via the project Performance Prediction for Dependable 6G Networks through Causal Artificial Intelligence (2024-02438).
This work is Co-funded by the European Union under Grant Agreement 101191936. Views and opinions expressed are however those of the author(s) only and do not necessarily reflect those of all SUSTAIN-6G consortium parties nor those of the European Union or the SNS-JU (granting authority). Neither the European Union nor the granting authority can be held responsible for them.


\bibliographystyle{IEEEtran}
\bibliography{ref}



\vfill\pagebreak

\end{document}